\documentclass[12pt]{article}
\usepackage{amsfonts,amsmath,amssymb}
\usepackage{amssymb}
\usepackage{epsfig}
\usepackage[rflt]{floatflt}

\begin{document}


\begin{center}
\vskip 3cm {\Large\bf On the First Order Phase Transitions Signal
in Multiple Production Processes}

\vskip 1cm J.Manjavidze and A.Sissakian\\

\vskip 0.5cm JINR, Dubna, Russia
\end{center}

\begin{abstract}
We offer the parameter, interpreted as the "chemical potential",
which is sensitive to the first order phase transition: it must
decrease with number of evaporating  (produced) particles (hadrons)
if the (interacting hadron or/and QCD plasma) medium is boiling and
it increase if no phase transition occur. The main part of the paper
is devoted to the question: how one can measure the "chemical
potential" in the hadron inelastic processes. Our definition of this
parameter is quite general but assumes that the hadron multiplicity
is sufficiently large. The simple transparent phenomenological
lattice gas model is considered for sake of clarity only.
\end{abstract}

\section{Introduction}

Despite the fact that the first order phase transition in the ion
collisions is widely discussed both from theoretical \cite{1} and
experimental \cite{2} points of view the feeling of some
dissatisfaction nevertheless remain. To all appearance the main
problem consists in absence of the single-meaning directly
measurable ("order") parameter(s) which may confirm this phenomenon
in the high energy experiment. Our aim is to offer such parameter,
explain its physical meaning and to show how it can be measured.

We guess that to observe first order phase transition it is
necessary to consider very high multiplicity (VHM) processes. Then
in this multiplicity region exist following parameter:
\begin{equation}
\mu(n,s) \simeq- \frac{T(n,s)}{n}\ln\sigma_n(s), \label{1.1}
\end{equation}
Here $\sigma_n$ is the normalized to unite multiplicity distribution
which can be considered in the VHM region as the "partition
function" of the $equilibrium$ system, see Appendix, and $T$ is the
mean energy of produced particles, i.e. $T$ is associated with
temperature. Continuing the analogy with thermodynamics one can say
that $(-T\ln\sigma_n)/n$ is the Gibbs free energy per one particle.
Then $\mu$ can be interpreted as the "chemical potential" measured
with help of $observed$ particles \footnote{Notice that one may
consider $n$ as the multiplicity in the experimentally observable
range of phase space.} in a free state.

We assume that the system obey the equilibrium condition, i.e.
produced particles energy distribution can be described with high
accuracy by Boltzmann exponent, or, it is the same, the inequality
(\r{b}) must be satisfied. This assumption defines the "VHM region"
\cite{physrep}~. It must be underlined that existence of the "good"
parameter $T$ does not assumes that the whole system is thermally
equilibrium, i.e. the energy spectrum of unobserved particles may be
arbitrary in our "inclusive" description.

The definition (\r{1.1}) is quiet general. It can be used both for
hadron-hadron and ion-ion collisions, both for low and high
energies. It is model free and operates only with "external"
directly measurable parameters. The single indispensable condition:
we work  in the $VHM$ region of observed particles. It is evident
that such generality has definite defect: measuring $\mu$ one can
not say something about details of the process.

This "defect" have following explanation. The point is that the
classical theory of phase transitions have dealing immediately with
the $internal$ properties of media in which the transition occur.
But in our, "$S$-matrix", case one can examine only the $external$
response on the phase transition in the form of created mass-shell
particles.

Continuing the analogy with the boiling, we are trying to define the
boiling by the number of evaporating particles. The effect is
evidently seen if the number of such particles is very large, i.e.
in the VHM case. The "order parameter" is the work needed for one
particle production, i.e. coincides with the "chemical potential".
In the boiled "two-phase" region the media is unstable against
"evaporation" of particles, i.e. the chemical potential must
decrease with number of produced particles.

We offer quantitative answers on the following three question.

(A) {\it In what case one} may {\it observe first order phase transition.}\\
We will argue that observation of VHM states are necessary to find
phase transition phenomenon. First of all the energy of produced
particles are small in VHM case. This means that the kinetic degrees
of freedom does not play essential role, i.e. they can not destroy,
wipe out, the phase transition phenomenon. The second reason is
connected with observation that in the VHM region one may use such
equilibrium thermodynamics parameters as the temperature $T(n,s)$,
chemical potential $\mu(n,s)$ and so on.

(B) {\it What we can measure.}\\ We will see that in VHM region
exist the estimation (\r{1.1}) where all quantities in r.h.s. are
measurables.

(C) {\it What kind effect one may expect.}\\  Chemical potential,
$\mu(n,s)$, by definition is the work which is necessary to
introduce, i.e. to produce, one particle into the system
\cite{isikhara}~. If the first order pase transition occurs then
$\mu(n,s)$ must decrease with $n$ in the two-phase ("boiling")
region. It is our general conclusion which will be explained using
lattice gas model.

\section{Definitions}

\subsection{}
We will start from simple generalization of well known formulae. Let
us consider the generating function
\begin{equation}
\rho(z,s)= \sum_{n=0}^\infty z^n \sigma_n(s).
\end{equation}
For sake of simplicity $\sigma_n$ is normalized so that
\begin{equation}
\rho(1,s)=1.\label{}
\end{equation}
One may use inverse Mellin transformation:
\begin{equation}
\sigma_n(s)=\frac{1}{2\pi i}\oint
\frac{dz}{z^{n+1}}\rho(z,s)\label{mel}
\end{equation}
to find $\sigma_n$ if $\rho(z,s)$ is known. Noting that $\sigma_n$
have sharp maximum over $n$ near mean multiplicity $\bar{n}(s)$ one
may calculate integral (\r{mel}) by saddle point method. The
equation:
\begin{equation}
n=\frac{\partial}{\partial\ln z}\ln\rho(z,s)\label{eq1}
\end{equation}
defines mostly essential value $z=z(n,s)$. Notice that
$\sigma_n\equiv0$ if the hadron multiplicity $n>n_{max}=\sqrt{s}/m$,
where $m$ is the characteristic mass of hadron. Production of
identical particles is considered for sake of simplicity. Therefore,
only $z<z_{max}=z(n_{max},s)$ have the physical meaning.

One may write $\rho(z,s)$ in the form:
\begin{equation}
\rho(z,s)=\exp\left\{ \sum_{l=0}^\infty z^l
b_l(s)\right\},\label{2.4}
\end{equation}
where the Mayer group coefficient $b_l$ can be
expressed through $k$-particle correlation function (binomial
moments) $c_k(s)$:
\begin{equation}
b_l(s)=\sum_{k=l}^\infty \frac{(-1)^{(k-l)}}{l!(k-l)!}c_k(s).
\label{}
\end{equation}
Let us assume now that in the sum:
\begin{equation}
\ln\rho_(z,s)=\sum_k \frac{(z-1)^k} {k!}c_k(s)\label{}
\end{equation}
one may leave first term. Then it is easy to see that
\begin{equation}
z(n,s)=n/c_1(s),~c_1(s)\equiv\bar{n}(s), \label{1.8}
\end{equation}
are essential and in the VHM region
\begin{equation}
\ln\sigma_n(s)=-n\ln\frac{n}{c_1(s)}(1+O(1/\ln n))=-n\ln
z(n,s)(1+O(1/\ln n)).\label{esti}
\end{equation}
Therefore, in considered case with
$c_k=0,~k>1,$ exist following asymptotic estimation for $n>>1$:
\begin{equation}
\ln\sigma_n\simeq-n\ln z(n,s),\label{1.10}
\end{equation}
i.e. $\sigma_n$ is defined in VHM region mainly by the solution of
Eq.(\r{eq1}) and the correction can not change this conclusion. It
will be shown that this kind of estimation is hold for arbitrary
asymptotics of $\sigma_n$.

If we understand $\sigma_n$ as the "partition function" in the VHM
region then $z$ is the $activity$ usually introduced in statistical
physics if the number of particles is not conserved. Correspondingly
the chemical potential $\mu$ is defined trough $z$:
\begin{equation}
\mu=T\ln z. \label{}
\end{equation}
Combining this definition with estimation (\r{1.10}) we define
$\sigma_n$ through $\mu$. But, if this estimation does not depend
from the asymptotics of $\sigma_n$, it can be used for definition of
$\mu(n,s)$ through $\sigma_n(s)$ and $T(n,s)$. Just this idea is
realized in (\r{1.1}).

\subsection{}
Now we will make the important step. To put in a good order our
intuition it is useful to consider $\rho(z,s)$ as the $nontrivial$
function of $z$. In statistical physics the thermodynamical limit is
considered for this purpose. In our case the finiteness of energy
$\sqrt{s}$ and of the hadron mass $m$ put obstacles on this way
since the system of produced particles necessarily belongs to the
energy-momentum surface\footnote{It must be noted that the canonical
thermodynamic system belongs to the energy-momentum shell because of
the energy exchange, i.e. interaction, with thermostat. The width of
the shell is defined by the temperature. But in particle physics
there is no thermostat and the physical system completely belongs to
the energy momentum surface.}. But we can continue theoretically
$\sigma_n$ to the range $n>n_{max}$ and consider $\rho(z,s)$ as the
nontrivial function of $z$.

Let us consider the analog generating function which has the first
$n<n_{max}$ coefficient of expansion over $z$ equal to $\sigma_n$
and higher coefficients for $n\geq n_{max}$ are deduced from
continuation of theoretical value of $\sigma_n$ to $n\geq n_{max}$.
Then the inverse Mellin transformation (\r{mel}) gives a good
estimation of $\sigma_n$ through this generating function if the
fluctuations near $z(n,s)$ are Gaussian or, it is the same, if
\begin{equation}
\left.\frac{|2n-z^3\partial^3\ln\rho(z,s)/\partial
z^3|}{|n+z^2\partial^2 \ln\rho(z,s)/\partial
z^2|^{3/2}}\right|_{z=z(n,s)}<<1. \label{z}
\end{equation}
Notice that if the estimation (\r{1.10}) is generally rightful then
one can easily find that l.h.s. of (\r{z}) is $\sim1/n^{1/2}$.
Therefore, one may consider $\rho(z,s)$ as the nontrivial function
of $z$ considering $z(n,s)<z_{max}$ if $n_{max}>>n>>1$.

Then it is easily deduce that the asymptotics of $\sigma_n(s)$ is
defined by the leftmost singularity, $z_c$, of in this way
generalized function $\rho(z,s)$ since, as it follows from
(\r{eq1}), the singularity "attracts" the solution $z(n,s)$ {\it in
the VHM region}. In result, we may classify asymptotics of
$\sigma_n$ in the VHM region if (\r{z}) is hold.

Thus our problem is reduced to the definition of possible location
of leftmost singularity of $\rho(z,s)$ over $z>0$. It must be
stressed that the character of singularity is not important for
definition of $\mu(n,s)$ in the VHM region at least with $O(1/\ln
n)$ accuracy. One may consider only three possibility: at
$n\to\infty$

(A) $z(n,s)\to z_c=1$,

(B) $z(n,s)\to z_c,~~1<z_c<\infty$,

(C) $z(n,s)\to z_c=\infty$.

Other possibilities are nonphysical or extremely rear.
Correspondingly one may consider only three type of asymptotics in
the VHM region:

(A) $\sigma_n>O(e^{-n})$. We will see that in this case the
isotropic momentum distribution must be observed, i.e. the energy,
$\varepsilon$, distribution in this case is Boltzmann-like, $\sim
e^{-\beta\epsilon}$;

(B) $\sigma_n=O(e^{-n})$. Such asymptotics is typical for hard
processes with large transverse momenta, like for jets \cite{jets}~;

(C) $\sigma_n<O(e^{-n})$. This asymptotic behavior is typical for
multiperipheral-like kinematics, where the longitudinal momenta of
produced particles are noticeably higher than the transverse ones
\cite{mpm}~.

We are forced to assume that the energy is sufficiently large, i.e.
$z_{max}$ is sufficiently close to $z_c$. In opposite case the
singularity would not be "seen" on experiment.

Our aim is to give physical interpretation of this three
asymptotics. The idea, as it follows from previous discussion, is
simple: one must explain the nature of singularity $z_c$. It must be
noted at the same time that in the equilibrium thermodynamics exist
only two possibility, (A) and (C) \cite{langer} and just the case
(A) corresponds to the first order phase transition.

Summarizing the results we conclude: {\it if the energy is
sufficiently large, i.e. if $z_{max}$ is sufficiently close to
$z_c$, if the multiplicity is sufficiently large, so that (\r{b}) is
satisfied and $z(n,s)$ can be sufficiently close to $z_c$, then one
may have confident answer on the question: exist or not first order
phase transition in hadron collisions.}

It must be noted here that the heavy ion collisions are the most
candidates since $z_c$ is easer distinguishable in this case.

\subsection{}
The temperature $T$ is the another problem. The temperature is
introduced usually using Kubo-Martin-Schwinger (KMS) periodic
boundary conditions \cite{niemi}~. But this way assumes from the
very beginning that the system (a) is equilibrium \cite{sch} and (b)
is surrounded by thermostat through which the temperature is
determined. The first condition (a) we take as the simplification
which gives the equilibrium state where the time ordering in the
particle production process is not important and therefore the time
may be excluded from consideration.

The second one (b) is the problem since there is no thermostat in
particle physics. For this reason we introduce the temperature as
the Lagrange multiplier $\beta=1/T$ of energy conservation law
\cite{physrep}~. In such approach the condition that the system is
in equilibrium with thermostat replaced by the condition that the
fluctuations in vicinity of $\beta$ are Gaussian.

The interesting for us $\rho(z,{s})$ we define through inverse
Laplace transform of $\rho(z,\beta)$:
\begin{equation}
\rho(z,s)=\int \frac{d\beta}{2\pi i\sqrt{s}} e^{\beta\sqrt{s}}
\rho(z,\beta). \label{}
\end{equation}
It is known \cite{sch} that if the interaction radii is finite then
the equation (of state):
\begin{equation}
\sqrt{s}=-\frac{\partial}{\partial\beta}\rho(z,\beta) \label{eq}
\end{equation}
have real positive solution $\beta(n,s)$ at $z=z(n,s)$. We will
assume that the fluctuations near $\beta(n,{s})$ are Gaussian. This
means that the inequality \cite{physrep}~:
\begin{equation}
\left.\frac{|\partial^3\ln \rho(z,\beta)/\partial\beta^3|}
{|\partial^2\ln\rho(z,\beta)/\partial\beta^2|^{3/2}}
\right|_{z=z(n,s),\beta=\beta(n,s)}<<1 \label{b}
\end{equation}
is satisfied. Therefore, we prepare the formalism to find
thermodynamic description of the processes of particle production
assuming that this $S$-matrix conditions of equilibrium (\r{z}) and
(\r{b}) are hold\footnote{Introduction of $\beta(n,s)$ allows to
describe the system of large number of degrees of freedom in terms
of single parameter, i.e. it is nothing but the useful trick. It is
no way for this reason to identify entirely $1/\beta(n,s)$ with
thermodynamic temperature where it has self-contained physical
sense.}.

We want to underline that our thermal equilibrium condition (\r{b})
have absolute meaning: if it is not satisfied then $\beta(n,s)$
loses every sense since the expansion in vicinity of $\beta(n,s)$
leads to the asymptotic series. In this case only the dynamical
description of $S$-matrix can be used.

It is not hard to see \cite{physrep} that
\begin{equation}
\frac{\partial^l}{\partial\beta^l}\ln R(z,\beta)
|_{z=z(n,s),\beta=\beta(n,s)}=<\prod_{i=1}^l(\epsilon_i-<\epsilon>)>_{n,s}
\label{}
\end{equation}
is the $l$-point energy correlator, where $<...>_{n,s}$ means
averaging over all events with given multiplicity and energy.
Therefore (\r{b}) means "relaxation of $l$-point correlations",
$l>2$, measured in units of the dispersion of energy fluctuations,
$l=2$. One can note here the difference of our definition of thermal
equilibrium from thermodynamical one \cite{bogol}~.

\subsection{}
Let us consider now the estimation (\r{1.1}). It follows from
(\r{mel}) that, up to the preexponential factor,
\begin{equation}
\ln\sigma(n,s)\approx-n\ln z(n,s)+ \ln\rho(z(n,s),s).\label{4.2}
\end{equation}
We want to show that, in a vide range of n from VHM
area,
\begin{equation}
n\ln z(n,s)\gtrsim \ln\rho(z(n,s),s).\label{}
\end{equation}
Let as consider now the mostly characteristic examples.

(A) {\it Singularity at $z=1$}.\\ This case will be considered in
Sec.3. In the used lattice gas approximation  $\ln{z}(n) \sim
n^{-5}$ and
\begin{equation}
\ln\sigma_n\approx n^{-4}=n\ln z(n)(1+O(1/n)).\label{}
\end{equation}
(B) {\it Singularity at $z=1+1/\bar{n}_j(s)<\infty$}:\\
$\ln\rho(z,s)=-\gamma\ln(1-\bar{n}_j(s)(z-1))$. In this case
$z(n)=1+1/\bar{n}_j(s)-1/n\bar{n}_j(s)$ and
$$\ln\sigma_n\approx -n/\bar{n}_j(s)+\gamma\ln n=$$
\begin{equation}
=-n/\bar{n}_j(s)(1+O(\ln n/n)).\label{}
\end{equation}
(C) {\it Singularity at $z=\infty$}: $\ln\rho(z,s)=c_k(s)(z-1)^k,~k\geq1$.\\
In this case $z(n)= (n/kc_k)^{1/k}>>1$ and
\begin{equation}
\ln\sigma_n\approx -n\ln z(n)(1+O(1/\ln n)).\label{}
\end{equation}
One can conclude:

(i) The definition (\r{1.1}) in the VHM region is rightful since the
correction falls down with $n$. On this stage we can give only the
estimation of correction but (\r{1.1}) gives the correct $n$
dependence.

(ii) Activity $z(n,s)$ tends to $z_c$ from the right in the case (A)
and from the left if we have the case(B) or (C).

(iii) The accuracy of estimation of the "chemical potential"
(\r{1.1}) increase from (C) to (A).

\section{Ising model: phase transition}

The physical meaning of singularity over $z$ \cite{lee} may be
illustrated by following simple model. As was mentioned above the
singularity at $z=1$ is interpreted as the first order phase
transition. Therefore, let us assume \cite{andreev} that $\beta$ is
so large that interacting particles strike together into clusters
(drops). Then the Mayer's group coefficient for the cluster from $l$
particle is
$$ b_l(\beta)\sim e^{-\beta\tau l^{(d-1)/d}},$$ where $\tau
l^{(d-1)/d}$, $l>>1$, is the surface tension energy, $d$ is the
dimension. Therefore, if $d>1$ the series over $l$ in (\r{2.4})
diverges at al $z>1$. At the same time, the sum (\r{2.4}) converge
for $z<1$.

We consider following analog model to describe condensation
phenomenon in the particle production processes. Let us cover the
space around interaction point by the net assuming that if the
particle hit the knot we have $(-1)$ and $(+1)$ in opposite case.
This "lattice gas" model \cite{isikhara} has a good description in
terms of Ising model \cite{newell}~. We may regulate number of down
oriented "spins", i.e. the number of produced particles, by external
magnetic field $\mathcal{H}$. Therefore the "activity"
$z=e^{-\beta\mathcal{H}}$, i.e. $-\mathcal{H}$ is the "chemical
potential" \cite{langer}~.

Calculation of the partition function means summation over all spin
configurations with constraint $\sigma^2=1$. Here the ergodic
hypothesis is used. It allows to exclude the time from
consideration.

To have the continuum model we may spread normally the
$\delta$-function of this constraint
\cite{wilson}~:$$\delta(\sigma^2-1)\sim
e^{-(1-\sigma^2)^2/\Delta}.$$ Therefore, the grand partition
function of the model in the continuum limit looks as follows
\cite{langer, kac}~:
\begin{equation}
\rho(\beta,z)=\int D\sigma e^{-S(\sigma,\mathcal{H})},\label{2.6}
\end{equation}
where the action
\begin{equation}
S(\sigma,\mathcal{H})=\int d^3x \left\{\frac{1}{2}(\nabla
\sigma)^2-\varepsilon\sigma^2+\alpha\sigma^4-\lambda\sigma\right\}.\label{}
\end{equation}
The structure of contributions in (\r{2.6}) essentially depends on
the sign of constant $\varepsilon$, see Fig.1 where the case
$\varepsilon>0$ is shown. Following notations was used:
\begin{equation}
\varepsilon\sim(1-\frac{\beta_c}{\beta}),~\alpha\sim\frac{\beta_c}{\beta}>0,
~\lambda\sim(\beta\beta_c)^{1/2}\mathcal{H},\label{2.8}
\end{equation}
where $1/\beta_c$ is the phase transition temperature. Phase
transition takes place if $\beta>\beta_c$ ($T<T_C$), i.e. we will
consider in present section $\varepsilon>0$. In this case the mean
spin $<\sigma>\neq0$. We will assume that $\beta>>\beta_c$ since in
this case the fluctuation around $<\sigma>$ are small and
calculations in this case became simpler. Considered model describes
decay of unstable vacuum \cite{coleman}~.

\begin{figure}[htbp]
\begin{center}
{\includegraphics[width=0.65\textwidth]{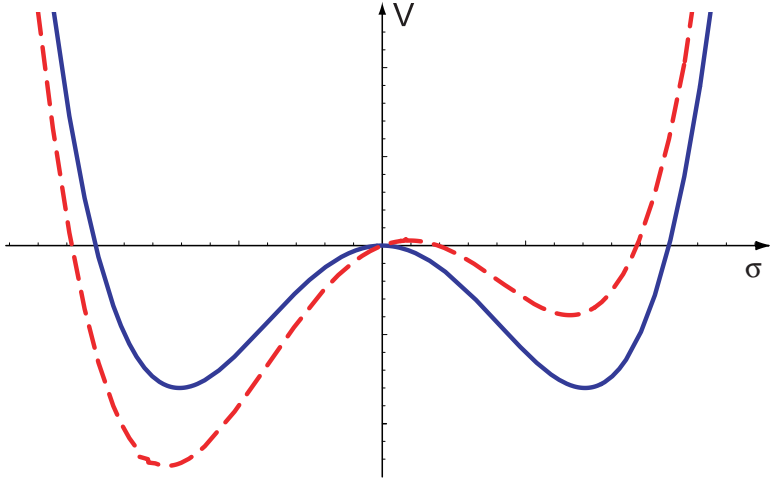}}
\caption{Solid line: undisturbed by $\mathcal{H}$ potential and
dotted line includes $\mathcal{H}$.} \label{fig1}
\end{center}
\end{figure}

The singularity over $\mathcal{H}$ appears by following reason. At
$\mathcal{H}=0$ the potential
\begin{equation}
v=-\varepsilon\sigma^2+\alpha\sigma^4\label{}
\end{equation}
have two degenerate minima at $\sigma=\pm
\sqrt{\varepsilon/2\alpha}$. The external field $\mathcal{H}<0$ we
destroy this degeneracy. But in this case the system in the
right-hand minimum (with the up-oriented spins) becomes unstable.

The branch point in the complex plane corresponds to this
instability. The discontinuity gives \cite{manj 1}~:
\begin{equation}
\rho(\beta,z)=\frac{a_1(\beta)}{\mathcal{H}^4}e^{-a_2(\beta)/\mathcal{H}^2},
\label{2.10}
\end{equation}
where ($\beta>\beta_c$) $$ a_1(\beta)=
\frac{\pi^2}{2}\left(\frac{8\beta R^4} {9\beta_c
A}\right)^{7/2}\left(\frac{1-\beta_c/\beta}{R^2}\right)^{3/4}\frac{R^4}{(\beta\beta_c)^2},
$$
\begin{equation}
a_2(\beta)=\frac{8\pi}{81\sqrt{2}}\frac{\beta}{\beta_c} \left(1-
\frac{\beta_c}{\beta}\right)^{7/2}\frac{R^4}{(A\beta_c^2)^2}.\label{}
\end{equation}
It must be noted that the eqs. (\r{eq1}) and (\r{eq}) have only one
solution: $$\beta(n,s)\to\beta_c,~~\ln z(n,s)\equiv l(n,s)\to0.$$ at
increasing $n$. This means that the singularities at $T=T_c$ and
$z=1$ attracts the solution:
\begin{equation}
l(n,\beta)\sim n^{-1/3}(\beta-\beta_c)^{7/6},
~~\beta(n,s)=\beta_c(1+\gamma/{n^4}),\label{2.12}
\end{equation}
where $\gamma$ is the positive constant.

In result,
\begin{equation}
\ln\rho_n(s)\sim -n^{2/3}(1/n^{2/3})^7\sim -1/n^4~~~(\sim -n\ln
z(n,s))\label{}
\end{equation}
decrees with $n$ and the chemical potential
\begin{equation}
\mu(n,s)\sim \frac{T_c}{n^5}(1+\gamma /n^4)^{-1}\label{}
\end{equation}
also decrees with $n$.

Some comments will be useful to this Section:

1. One may note that $\sigma_n$ is defined by the discontinuity the
the branch point in complex plane of $\ln z$ and decay of the
meta-stable states does not play any role.

\begin{figure}[htbp]
\begin{center}
{\includegraphics[width=0.65\textwidth]{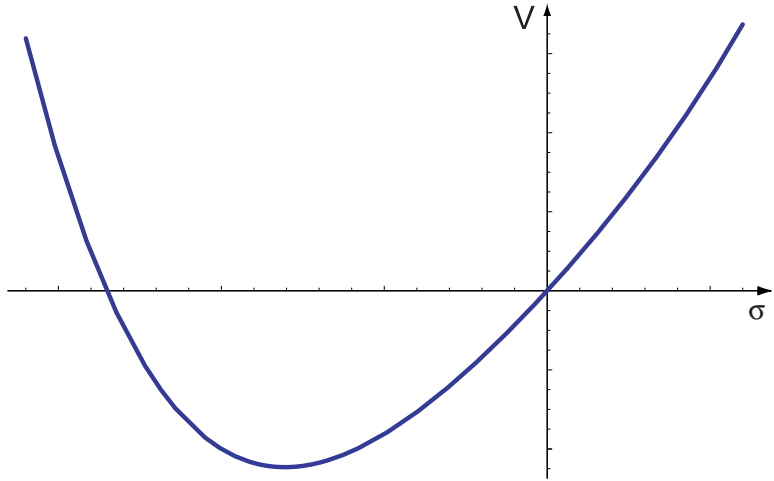}}
\caption{Stable ground state disturbed by $\mathcal{H}$.}
\label{fig2}
\end{center}
\end{figure}

2. It follows from (\r{2.12}) that, at fixed $\beta$,
\begin{equation}
\ln z_c\sim (1/n)^{1/3} <<1. \label{}
\end{equation}
This means that for
large $n$ our calculations are correct. At the same time, in VHM
region $z$ near unite is essential and it $decrees$ with $n$.

3. The work which is needed for production of one particle is $\sim
\ln z(n,s)/\beta(n,s)$. Therefore production of large number of
particles needs less work per one particle.

This conclusion have simple physical meaning  (see beginning of
present section). Let us consider decay of nonstable phase. The
decay happens through production of clusters (domains with down
oriented spins). The volume energy of cluster is $\propto R_0^3$,
where $R_0$ is radius of cluster. It burst the dimension of cluster.
If $R_0<R_c$, where $R_c$ is critical dimension of cluster, then the
formation of such cluster is improbable. But if $R_0>R_c$ then the
probability grows with radii of cluster. Latter explains why the
chemical potential falls down with multiplicity.

4. In the VHM region the temperature, $T(n.s)$, tends to its
critical value, $T_c$, and slowly depends on $n$.

\section{Ising model: stable minimum}

Let as consider the system with stable vacuum, $\beta < \beta_c$
($T>T_c$) in (\r{2.8}). In this case, see Fig.2, the potential
$v(\sigma)$ has unique minimum at $\sigma=0$. Switching on external
field $\mathcal{H}$ the minimum move and the average spin appears,
$\bar \sigma(\mathcal{H})\neq0$. One can find it from the equation:
$$-\triangle\sigma+2\varepsilon\sigma+4\alpha\sigma^3=\lambda,~ \varepsilon>0.$$ Having $\bar\sigma\neq0$ we
must expand the integral (\r{2.6}) near $\bar\sigma$:
\begin{equation}
\rho(\beta,z)=e^{\int
dx\lambda\bar\sigma}e^{-W(\bar\sigma)},\label{}
\end{equation}
where $W(\bar\sigma)$ expandable over $\bar\sigma$:
\begin{equation}
W(\bar\sigma)=\sum_{l=1}^ \infty \frac{1}{l}\int \prod_{i=1}^l\{dx_i
\bar\sigma(x_i;\mathcal{H})\}B_l(x_1, ...,x_l),\label{3.2}
\end{equation}
where $B_l$ is the $l$-point one particle irreducible vertex
function. In another wards, $B_l$ play the role of virial
coefficient. Comparing (\r{3.2}) with (\r{2.4}) one may consider
$\bar\sigma$ as the affective activity of group of $l$ particles.

The representation (\r{3.2}) is useful since in the VHM region the
density of particles is large and the particles momentum is small.
Then, remembering that the virial decomposition is equivalent of
decomposition over specific volume, calculating $B_l$ one may not go
beyond the one-loop approximation, i.e. we may restricted by the
semiclassical approximation.

Therefore, having large density one may neglect the spacial
fluctuations. In this case the integral (\r{2.6}) is reduced down to
the the usual Cauchy integral:
\begin{equation}
\rho(\beta,z)=\int_{-\infty}^{+ \infty} d\sigma
e^{-(\epsilon\sigma^2+\alpha\sigma^4+\lambda\sigma)}.\label{}
\end{equation}
In the VHM region $\lambda\sim\mathcal{H}\sim\ln z>>1$ is essential.
It is easy to see that
\begin{equation}
\bar\sigma\simeq -(\lambda/4\alpha)^{1/3}\label{}
\end{equation}
is the extremum. The estimation of integral near this $\bar\sigma$
looks as follows:
\begin{equation}
\rho(\beta,z)\propto\left\{12\alpha\left(\frac{\lambda}{4\alpha}\right)^{2/3}\right\}^{-1/2}
e^{3\lambda^{4/3}/4(4\alpha)^{1/3}}.\label{}
\end{equation}
This leads to increasing with $n$ activity:
\begin{equation} l(n,s) \sim
n^{8/3}\label{}
\end{equation}
and
\begin{equation}
\beta(n,s) \sim n^{2/3}.\label{}
\end{equation}
In result,
\begin{equation}
\ln \rho_n(s)\sim - n^{11/3}~(\sim -n\ln{z(n,s)}). \label{}
\end{equation}
A few comments at the end of this section:

(i) Cross section falls dawn in considered case faster then
$O(e^{-n})$. The estimation: $$ \rho_n(s)\sim
e^{-n\ln\bar{z}(n,s)}$$ gives the right expression in the VHM
region.

(ii) The chemical potential increase with $n$:
\begin{equation}
\mu(n,s)=- (T(n,s)/n)\ln\sigma_n(s)\sim n^{2}.\label{}
\end{equation}

\section{Conclusions}
We may conclude that:

(i) We found definition of chemical potential (\r{1.1}). This
important observable can be measured on the experiment directly
where $T(n,s)$ is the mean energy of produced particles at given
multiplicity $n$ and energy $\sqrt s$.

((ii) Being in the VHM region one may consider that
$\mu(n,s)=O(1/n)$ at comparatively high multiplicities and it rise,
$\mu(n,s)=O(n)$ with rising multiplicity, Fig.3, at comparatively
low multiplicities. The transition region is defines the critical
temperature $T_c$. But it is quiet possible that the condition
(\r{b}) allows to see only one branch of the curve shown on Fig.3.

iv) The simplest example of finite $z_c$ presents the jet considered
in the case (B), Fig.1. Hence case (C) has pure dynamical basis and
can not be explained by equilibrium thermodynamics.

\begin{figure}[h]
\begin{center}
{\includegraphics[width=0.65\textwidth]{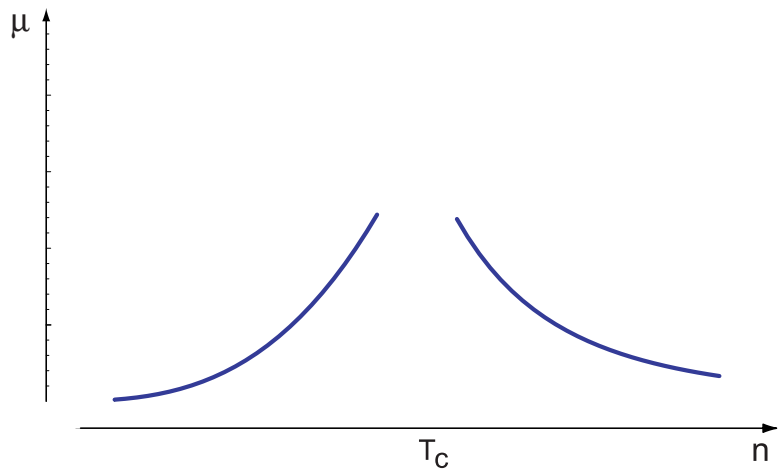}}
\caption{Chemical potential $\mu(n,s)$ as the function of
multiplicity $n$. $T_c$ is the critical temperature. Breakthrough
is the "two-phase" region.} \label{fig3}
\end{center}
\end{figure}

\section*{Acknowledgements}
We would like to thank participants of 7-th International Workshop
on the "Very High Multiplicity Physics" (JINR, Dubna) for
stimulating discussions.

\end{document}